\pdfoutput=1
\documentclass[letterpaper,10pt]{article}
\usepackage{osameet2}

\usepackage{amsmath,amssymb}
\usepackage{graphicx}
\usepackage{epstopdf}
\usepackage[pdftex,colorlinks=true,bookmarks=false,citecolor=blue,urlcolor=blue]{hyperref} 

\begin{document}

\title{Optimized aperiodic broadband thermal emitters\\
 for use as light bulb filaments}

\author{Christopher H. Granier,$^1$ Sim\'{o}n G. Lorenzo,$^{1,2,*}$ Chenglong You,$^1$ Georgios Veronis,$^{2,3}$ Jonathan P. Dowling$^1$}
\address{[1]Hearne Institute for Theoretical Physics and Department of Physics and Astronomy, Louisiana State University, Baton Rouge, Louisiana 70803\\\relax
[2]Center for Computation and Technology, Louisiana State University, Baton Rouge, Louisiana 70803\\\relax
[3]School of Electrical Engineering and Computer Science, Louisiana State University, Baton Rouge, Louisiana 70803}
\email{slorenzo6314@gmail.com}

\begin{abstract}
We present optimized aperiodic structures for use as broadband thermal incandescent emitters which are capable of increasing the emittance by nearly a factor of two over the visible wavelength range when compared to bulk tungsten. These aperiodic multilayer structures are designed with alternating layers of tungsten and air or tungsten and silicon carbide on top of a tungsten substrate. We investigate the properties of these structures for use as lightbulb filaments. We find that these structures greatly enhance the emittance over the visible wavelength range, while also increasing the overall efficiency of the bulb and could lead to a decrease in incandescent lightbulb power consumption by nearly 50$\%$.
\end{abstract}


\section{Introduction}

Conventional incandescent lightbulbs are composed of a tungsten filament inside a bulb which is filled with inert gas. These devices are less than 10$\%$  efficient \cite{solar,solar2} due to the fact that most of the applied power is radiated as infrared radiation and not as visible light. There has been much recent work in an attempt to improve the efficiency of household and commercial lighting. Compact fluorescent lamps (CFLs), improve efficiency slightly, but have issues with slow turn on time, higher cost, degradation of performance over CFL lifetime, and most significantly, these bulbs contain mercury and pose an environmental hazard if broken or disposed of in a landfill. The most efficient lighting commercially available is the light-emitting diode (LED) which possesses efficiencies between 25 and 35$\%$. LEDs, however, are extremely expensive (due to fabrication costs) which puts them out of reach of users in emergent countries. Hence, incandescent bulbs still account for the vast majority of home lighting in developing countries.

In this paper, we design aperiodic broadband thermal emitters in the visible wavelength range, composed of alternating layers of tungsten and air or tungsten and silicon carbide above a semi-infinite tungsten substrate, for use as lightbulb filaments \cite{solar,SPIELightbulb,mit}. We use a hybrid optimization method, consisting of a micro-genetic global optimization algorithm coupled to a local optimization algorithm, to maximize the power emitted by the structures in the visible. We then maximize the efficiency of the structures without significantly decreasing the power emitted in the visible. We find that the optimized structures obtained using this process show drastically improved emittance over the visible wavelength range when compared to bulk tungsten. The emittance of the optimized structures approaches the one of a perfect blackbody. Thus, these structures would lead to a great decrease in incandescent lightbulb power consumption. We also find that the optimized tungsten-air structure is essentially equivalent to a reduced structure consisting of a thin tungsten layer and an air layer above a thick tungsten substrate. Similarly, we find that the optimized tungsten-silicon carbide structure is essentially equivalent to a reduced structure which consists of a single silicon carbide layer above a thick tungsten substrate. 

The remainder of the paper is organized as follows. In Section II we discuss the computational techniques used. The results obtained using these techniques are presented in Section III. Finally, our conclusions are summarized in Section IV.

\section{Theory}

We model a structure composed of infinite slabs of material of varying aperiodic thicknesses above a semi-infinite tungsten substrate as depicted in Fig. 1. Utilizing the transfer matrix method \cite{matrixxfer}, we calculate the transmittance, reflectance, and absorptance of the structure for both TE and TM polarized light. Light is incident from air at an angle $ \theta $ to the structure. We make use of experimental data for the wavelength-dependent indices of refraction, both real and imaginary parts, for silicon carbide and tungsten \cite{indices} in all calculations done in this paper.  We choose silicon carbide and tungsten due to their high melting point, which is necessary for thermal emitter applications. Since the tungsten substrate is taken to be semi-infinite, the transmittance is identically zero, so that:
\begin{equation}
A_{\rm TE/TM}(\lambda,\theta)=1-R_{\rm TE/TM}(\lambda,\theta),
\end{equation}
where $ A_{\rm TE/TM} $ is the TE/TM absorptance, $ R_{\rm TE/TM} $ is the TE/TM reflectance, and $\lambda$ is the wavelength. While we only calculate absorptance, reflectance, and, in principle, transmittance, we make use of Kirchhoff's second law and conservation of energy to equate absorptance ($A_{\rm TE/TM}$) and emittance ($\epsilon_{\rm TE/TM}$) under thermal equilibrium \cite{kirchoff}. 
\begin{figure}[h]
  \centering
  \includegraphics[width=6.5cm]{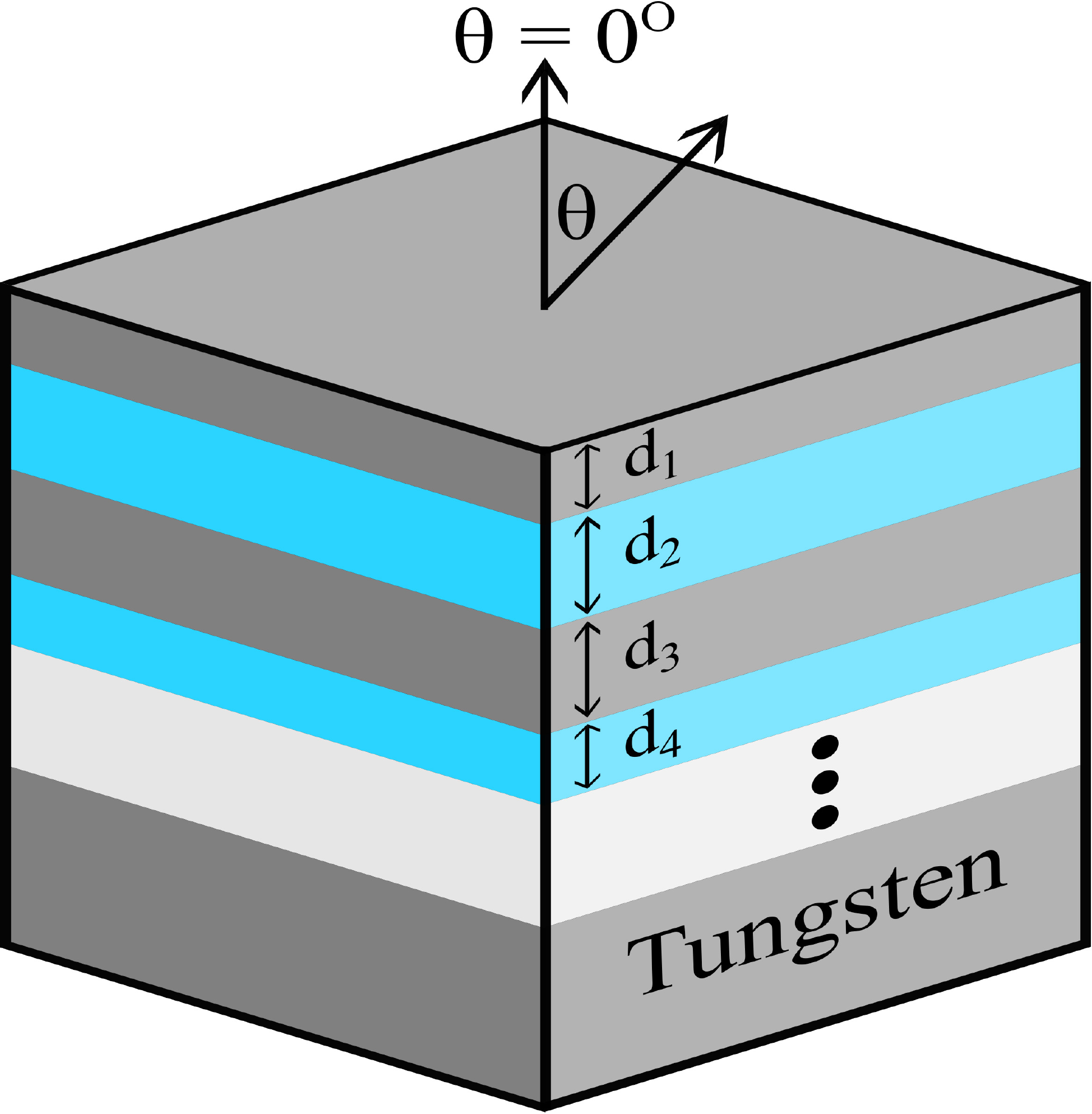}\\
  \caption{Schematic of the structure optimized. It consists of alternating layers of tungsten and air or tungsten and silicon carbide above a semi-infinite tungsten substrate.}\label{1}
  
\end{figure}
More specifically, the power radiated per unit area and wavelength is given by the Planck blackbody spectrum (in units of Watt per square meter per nanometer): 
\begin{equation}
B(\lambda,T)=\frac{2hc^2}{\lambda^5}(e^\frac{hc}{\lambda k_{B}T}-1)^{-1},
\end{equation}
 where $h$ is Planck's constant, $c$ is the speed of light, $\lambda$ is the wavelength, $T$ is the temperature, and $k_{B}$ is the Boltzmann constant. We define the normalized power radiated per unit area and unit wavelength by a given structure, $\bar{\mu}(\lambda,\theta) $ as the ratio of the power emitted per unit area and wavelength by the structure to the maximum emitted by a blackbody
 \begin{equation}
 	\bar{\mu}(\lambda,\theta)= \frac{\epsilon_{\rm Total}(\lambda,\theta)B(\lambda,T=2700K)}{\underset{\lambda}\max[B(\lambda,T=2700K)]}.
 \end{equation}
 Here, $ \epsilon_{\rm Total}(\lambda,\theta)=[ \epsilon_{\rm TE}(\lambda,\theta)+\epsilon_{\rm TM}(\lambda,\theta)]/2$. We choose an operating temperature of 2700 K, below the melting point of silicon carbide and tungsten and near the operating temperature of modern incandescent light bulbs \cite{physofeveryday}.
 
  Conventional incandescent light bulbs are made by drawing tungsten into a thin wire and running a current through the wire. This process, via the resistance of the wire, heats the wire to a temperature between 2500 K and 3000 K and produces light over broad wavelength range. Due to the operating temperature of the filament as well as the emittance spectra of bulk tungsten, this process is inefficient and produces both a large amount of waste heat and light which is not in the visible wavelength range  \cite{chemicalcomp,envchem}. 
  
 \begin{figure}[h]
   \centering
   \includegraphics[width=9cm]{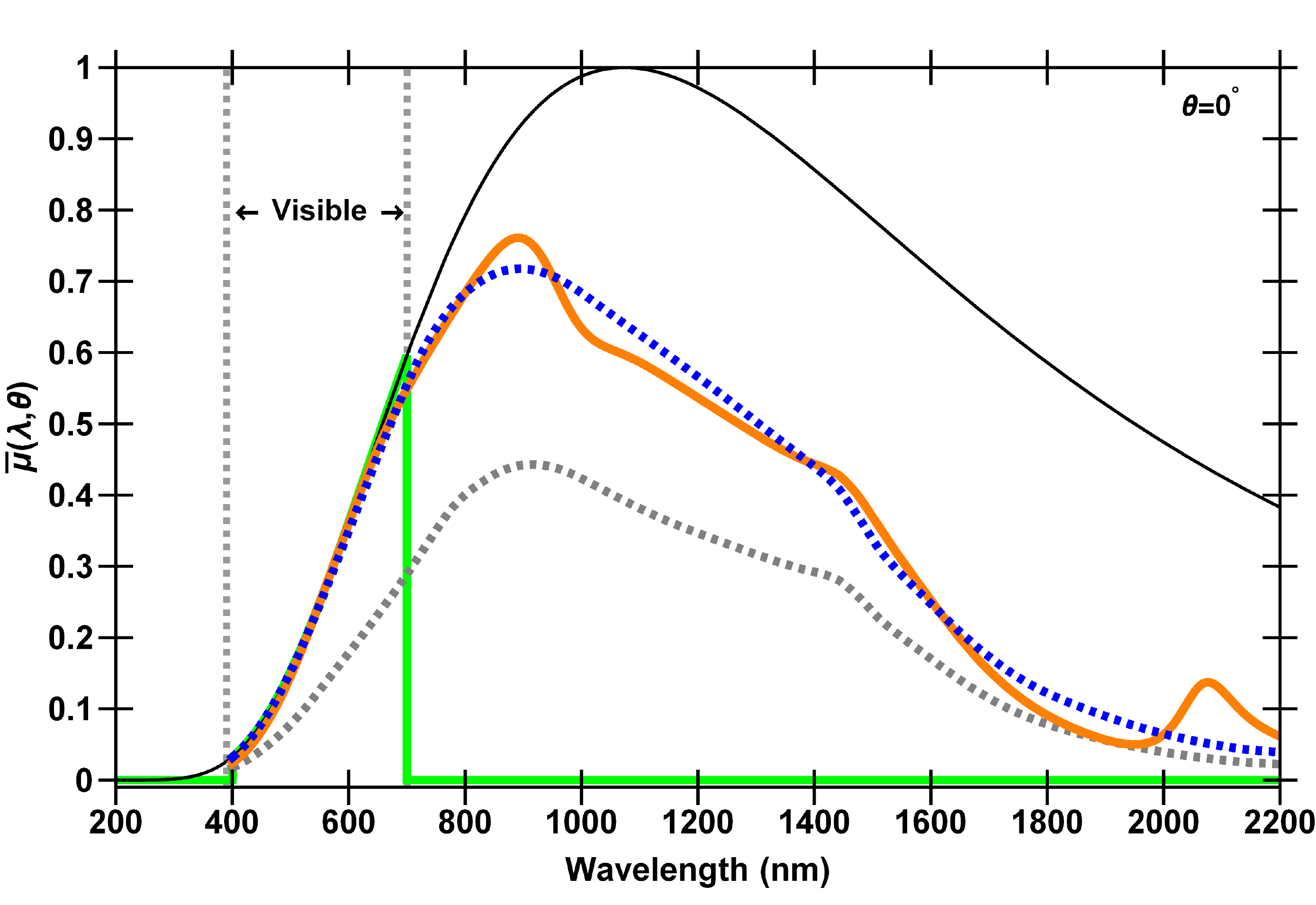}\\
   \caption{Normalized power emitted per unit area and unit wavelength, in the normal direction, $\bar{\mu}(\lambda,\theta=0^\circ)$, as a function of wavelength for the optimized structures compared to both a perfect blackbody at the same temperature as well as a bulk tungsten filament. The solid black line depicts a perfect blackbody at 2700 K. The solid green line shows the performance of an ideal filament operating at 2700 K. The dashed gray line shows the performance of a bulk tungsten filament. The solid orange and dashed blue lines show the performance of the optimized TSiC and TA structures, respectively. The layer thicknesses of the optimized structures, in units of nanometers, beginning with the tungsten layer bordering air are: \{3.4, 90, 163, 500\} for the tungsten-air structure and \{0, 39, 66, 319\} for the TSiC structure.  }\label{1}
 \end{figure}
 \begin{figure*}[!ht]
   \centering
   \includegraphics[width=16cm]{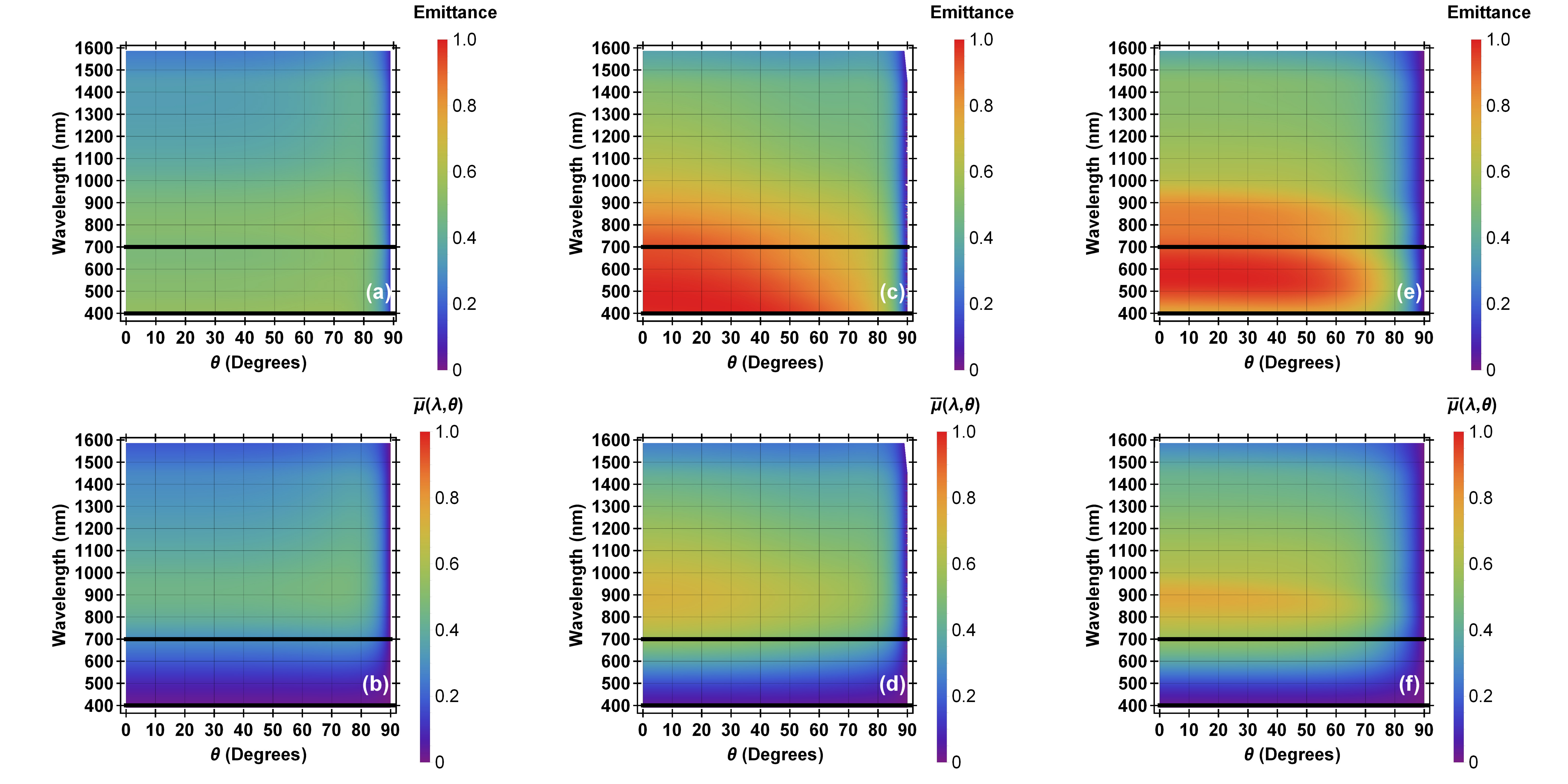}\\
   \caption{(a, b) Emittance and normalized power emitted per unit area and unit wavelength $\bar{\mu}(\lambda,\theta)$ of our reference bulk tungsten filament as a function of wavelength and angle. (c, d) Emittance and normalized power emitted per unit area and unit wavelength $\bar{\mu}(\lambda,\theta)$ for the optimized TA structure as a function of wavelength and angle. (e, f) Emittance and normalized power emitted per unit area and unit wavelength $\bar{\mu}(\lambda,\theta)$ for the optimized TSiC structure as a function of wavelength and angle. The layer thicknesses of the optimized TA and TSiC structures are given in Fig. 2. The horizontal, thick, black lines show the visible wavelength range.}\label{1}
 \end{figure*}

In traditional bulk tungsten filaments, in excess of  $90\%$ of the power which enters the filament is released as infrared heat radiation. Additionally, of the approximate $10\%$ of the power which is released as radiation, only a small percentage of the light is emitted in the visible range (Fig. 2). We define the efficiency of a given structure as the ratio of the normalized power emitted in the visible wavelength range ($\lambda_{1} = 400\  \rm nm $ to  $\lambda_{2} = 700\  \rm nm$) in the normal direction to the total power emitted:
   \begin{equation}
   	\eta=\frac{\int\limits_{\lambda_1}^{\lambda_2}	\bar{\mu}(\lambda,\theta = 0^\circ)\,d\lambda\\}{\int\limits_{0}^{\infty}	\bar{\mu}(\lambda,\theta = 0^\circ)\,d\lambda\\}.
   \end{equation}
 
Our reference structure is a bulk tungsten filament; in other words, the structure in Fig. 1 without any layers on top of the substrate is used for comparison. We envision growing the multilayers on the cylindrical tungsten filament by vapor deposition or molecular beam epitaxy. So long as the radius of curvature of the substrate is much larger than the layer thicknesses, the one-dimensional approximation is justified. Using Eq. (4), we find that for a bulk tungsten filament operating at $T= 2700 $ K, only $\sim 11.55 \%$ of the power radiated  lies in the visible wavelength range. The combination of these effects leads to an extremely low efficiency for conventional incandescent tungsten filament light bulbs.  
 
Another factor which contributes to the overall low efficiency of conventional tungsten filaments is the fact that tungsten is highly reflective over the visible wavelength range. Hence, tungsten is a relatively poor emitter, with an average emittance over the visible spectrum of approximately $ 50\%$. We define the enhancement factor \textit{Q} of a given structure compared to our reference structure as the ratio of the total power emitted in the visible in the normal direction by the structure at $T= 2700 $ K to the power emitted by our reference bulk tungsten filament heated to the same temperature:

 \begin{equation}
 	Q=\frac{\int\limits_{\lambda_1}^{\lambda_2}	\bar{\mu}(\lambda,\theta = 0^\circ)\,d\lambda\\}{\int\limits_{\lambda_1}^{\lambda_2}	\bar{\mu}_{\rm Bulk}(\lambda,\theta = 0^\circ)\,d\lambda\\}.
 \end{equation}

\section{Results}
We are interested in finding structures which exhibit a plateau-like emittance over the visible wavelength range. That is, we seek a structure which possesses $\epsilon = 1$ in the visible wavelength range ($\lambda_{1} \le \lambda \le \  \lambda_{2} $) and $\epsilon = 0$ elsewhere (corresponding to the solid green line in Fig. 2) in an effort to maximize the power emitted by the structure in the visible while maximizing its efficiency.

We use a hybrid optimization method consisting of a micro-genetic global optimization algorithm \cite{genetic1,genetic0,genetic2,genetic3,genetic4,Granier1,Granier2}, coupled to a local optimization algorithm \cite{NLOPT1,NLOPT2}, to determine the best dimensions for four-layer structures. The genetic algorithm is an iterative optimization procedure which starts with a randomly selected population of potential solutions and evolves toward improved solutions; once the population converges, the local optimization algorithm finds the local optimum. The process retains the best structure found and is iteratively repeated.

Specifically, we first maximize the fitness function $ F_{1} $ where
\begin{equation} F_{1}=\int\limits_{\lambda_1}^{\lambda_2}\bar{\mu}(\lambda,\theta = 0^\circ)\,d\lambda\\.
\end{equation}
This process is carried out to enhance the emittance in the visible when compared to that of our reference bulk tungsten filament.

We then seek to maximize the efficiency of the structure [Eq. (4)] without significantly decreasing the enhancement of the power emitted in the visible compared to bulk tungsten [Eq. (5)]. Due to the operating temperature of the structure ($ T = 2700 $ K), any radiation occurring at $\lambda \le 400 \rm nm $ can be treated as negligible. Thus, we minimize the normalized power emitted by the structure for $\lambda \ge 700 \rm \ nm $, $F_{2}, $
\begin{equation}
F_{2}=\int\limits_{\lambda_2}^{\infty}	\bar{\mu}(\lambda,\theta = 0^\circ)\,d\lambda\\,
\end{equation}
subject to the constraint that the fitness function $F_1$ for the structure is at least 98 \% of the previously calculated maximum $F_1$.

We consider structures of four layers composed of alternating layers of tungsten and air (henceforth referred to as the TA structures) or alternating layers of tungsten and silicon carbide (henceforth referred to as the TSiC structures) above a semi-infinite tungsten substrate (Fig. 1). We found that using more than four layers leads to negligible improvement to the performance of the structures. We optimize the emittance of the structures using the process outlined above [Eqs. (6) and (7)].
 \begin{figure}[!ht]
 \centering
   \includegraphics[width=8.0cm]{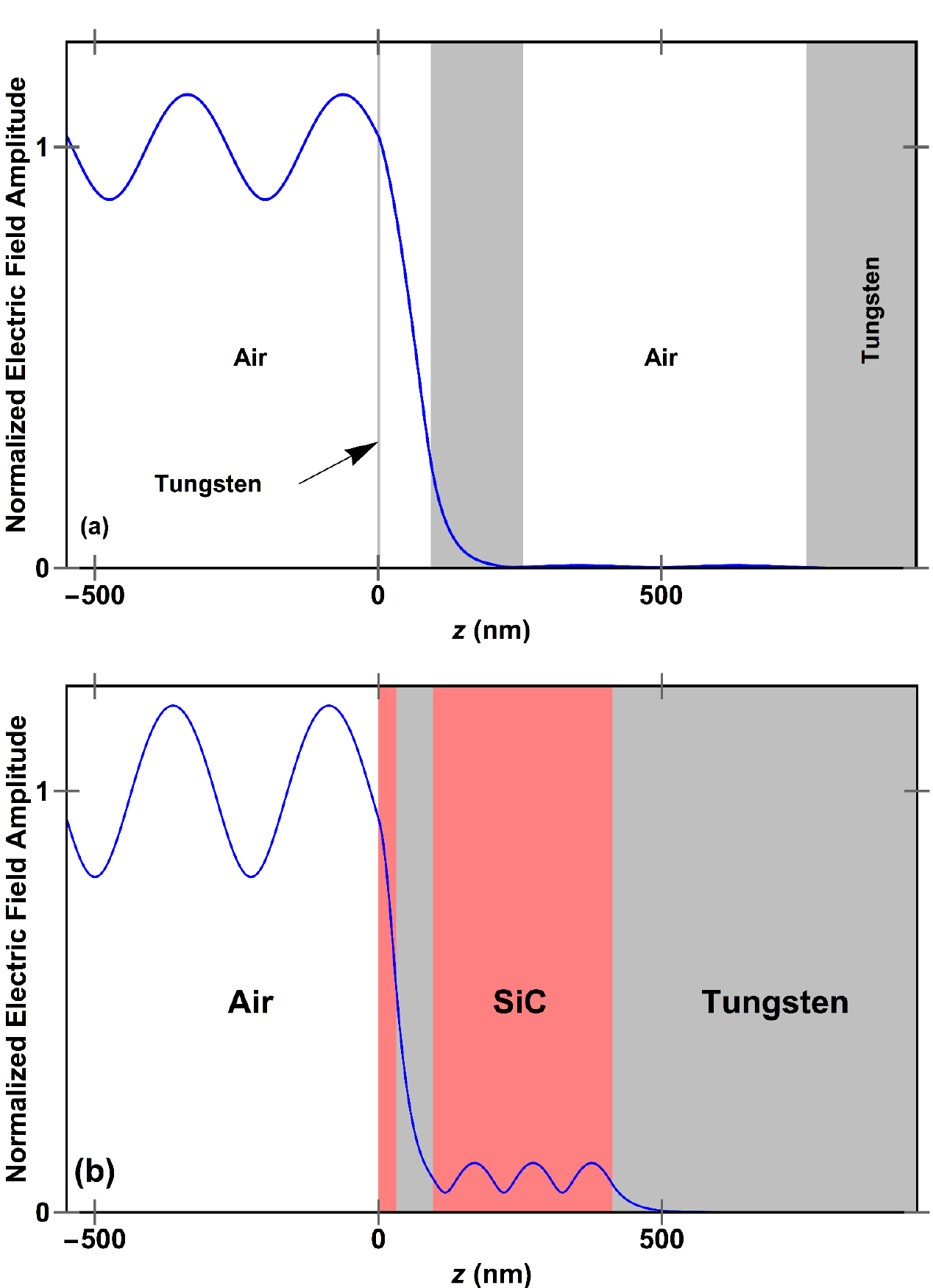}\\
   \caption{Profile of the electric field amplitude, normalized with respect to the field amplitude of the incident plane wave for the TA and TSiC optimized structures described in Fig. 2. The structures are excited by a normally incident plane wave at the wavelength of  $\lambda_{0}$ = $550$ nm, near the center of the visible wavelength range.
 (a) For the TA structure, the ratio of the power absorbed inside each layer to the total power absorbed in the structure was calculated and from left to right, beginning with tungsten, is: \{0.79, 0, 0.21, 0, 0\}. That is, $\sim$79\% of the power is absorbed in the first tungsten layer adjacent to air. (b) For the TSiC aperiodic structure, the thickness of the first tungsten layer is zero; thus, the layer adjacent to air is silicon carbide. The ratio of the power absorbed inside each layer to the total power absorbed in the structure from left to right, beginning with SiC, is: \{0, 0.99, 0, 0.01\}. That is, $\sim$99\% of the power is absorbed in the first tungsten layer.}
   \label{1}
 
 \end{figure}
 
Both the TA and TSiC optimized structures that we obtained using this process show drastically improved emittance (by nearly a factor of two) over the visible wavelength range when compared to bulk tungsten (Fig. 2). Thus, these structures could lead to a decrease in incandescent light bulb power consumption by nearly 50 \%. More specifically, a light bulb consuming $30 \rm \ W$ of power using our design would radiate approximately the same energy in the visible as a conventional incandescent bulb which consumed $60 \rm \ W$ of power. The emittance of the optimized structures approaches the one of a perfect blackbody. By comparing the bulk tungsten's emittance and normalized power emitted per unit area and unit wavelength $\bar{\mu}$ [Figs. 3(a) \& 3(b)] to that of both TA [Figs. 3(c) \& 3(d)] and TSiC structures [Figs. 3(e) \& 3(f)], we observe that the optimized structures provide greatly enhanced emittance over the visible wavelength range when compared to bulk tungsten. As a result, the optimized structures also greatly enhance the power emitted by the structures in the visible. A detailed comparison of the enhancement factor \textit{Q} and the efficiency $\eta$ of the different structures can be found in Table 1.

\begin{table}[!ht]
	\begin{center}
		\caption{Comparison of the enhancement factor $\textit{Q}$ and the efficiency $\eta$ using a bulk tungsten filament as a reference point, of the optimized structures described in Fig. 2.}
		\begin{tabular}{ | l | l | l | l |}
			\hline
			Structure & $Q$ & $\eta$  \\ \hline
			Tungsten & $1.0$ & $11.55 \%$\\ \hline
			TA & $1.949$ & $13.62\%$ \\ \hline
			Reduced TA & $1.949$ & $13.61 \%$\\ \hline
			TSiC & $1.950$ & $13.77 \%$\\\hline
			Reduced TSiC & $1.943$ & $13.54 \%$\\
			\hline
		\end{tabular}
	\end{center}
\end{table}
In Fig. 4(a), we show the profile of the electric field amplitude normalized with respect to the field amplitude of the incident plane wave for the optimized TA structure. The structure is excited by a normally incident plane wave at the center of the visible spectrum, $ \lambda = 550\  \rm nm$. The TA structure is nearly perfectly impedance-matched to air, as only $\sim 10\%$ of the incident power is reflected. Additionally, there is no field enhancement in the structure. Therefore, we conclude that the high emittance is not associated with any strong resonance, which is consistent with the broad-angle and broadband emittance of the structure [Fig. 3(c)].

We also consider the effect of individual layer thicknesses on the emittance spectra of the TA structures [Figs. 5(a) \& 5(b)]. The optimized thickness of the first tungsten layer adjacent to air is found to be 3.4 nm due to the fact that a thinner tungsten layer does not achieve near-blackbody emittance in the visible wavelength range; while, a thicker tungsten layer decreases the efficiency by emitting more in the near infrared [Fig. 5(a)]. A similar phenomenon is found with the selection of the 90 nm air gap. At a given wavelength, the emittance as a function of the air layer thickness exhibits peaks associated with the Fabry-Perot resonances of the structure [Fig. 5(b)]. Again, this particular thickness is selected due to the fact that a structure having a smaller air gap possesses smaller emittance in the visible; while, a structure having a larger air gap decreases the efficiency by emitting more in the infrared [Fig. 5(b)]. We note that the enhancement factor $Q$ as a function of the air gap thickness exhibits additional peaks. However, the maximum enhancement is obtained at the first peak, since for higher order peaks the emittance is not high in the entire visible wavelength range [Fig. 5(b)]. 
 \begin{figure*}[!ht]
   \centering
   \includegraphics[width=16cm]{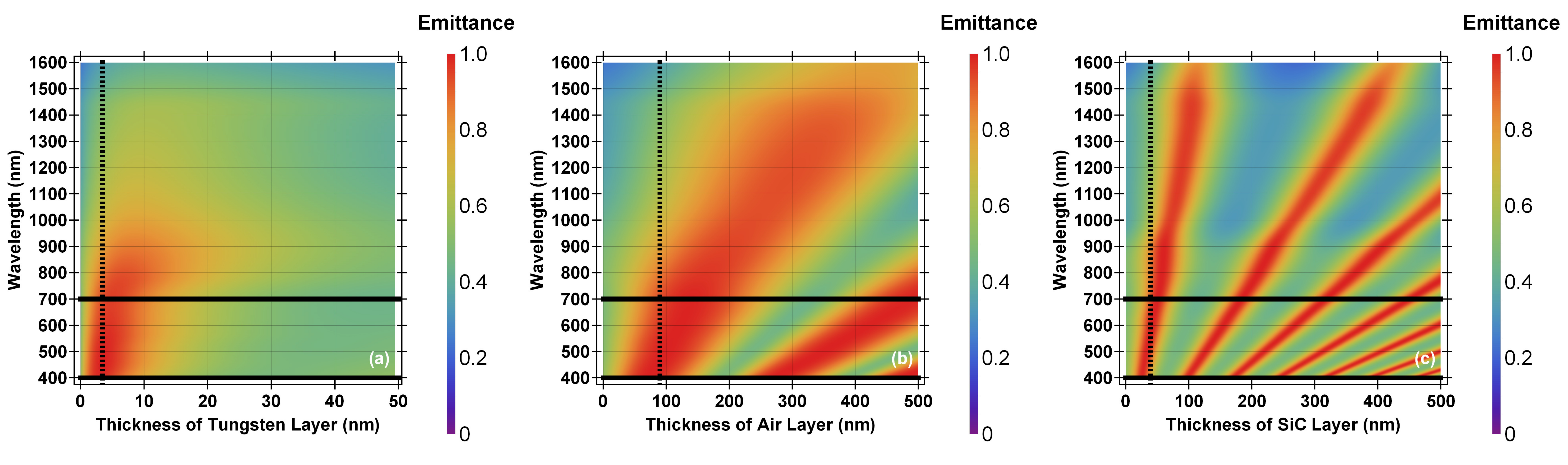}\\
   \caption{(a) Emittance in the normal direction of a four-layer TA structure (Fig. 1) as a function of wavelength and the thickness of the first tungsten layer adjacent to air. All other layer thicknesses are as in the optimized TA structure described in Fig. 2. (b) Emittance in the normal direction of a four-layer TA structure (Fig. 1) as a function of wavelength and the thickness of the air layer below the first tungsten layer. All other layer thicknesses are as in the optimized TA structure described in Fig. 2. (c) Emittance in the normal direction of a three-layer TSiC structure (Fig. 1) as a function of wavelength and the thickness of the first silicon carbide layer adjacent to air. All other layer thicknesses are as in the optimized TSiC structure described in Fig. 2. The horizontal, thick, black lines show the visible wavelength range; while, the dashed vertical line denotes the optimized structure dimension. }\label{1}
 \end{figure*}

 \begin{figure}[!ht]
 \centering
   \includegraphics[width=8.0cm]{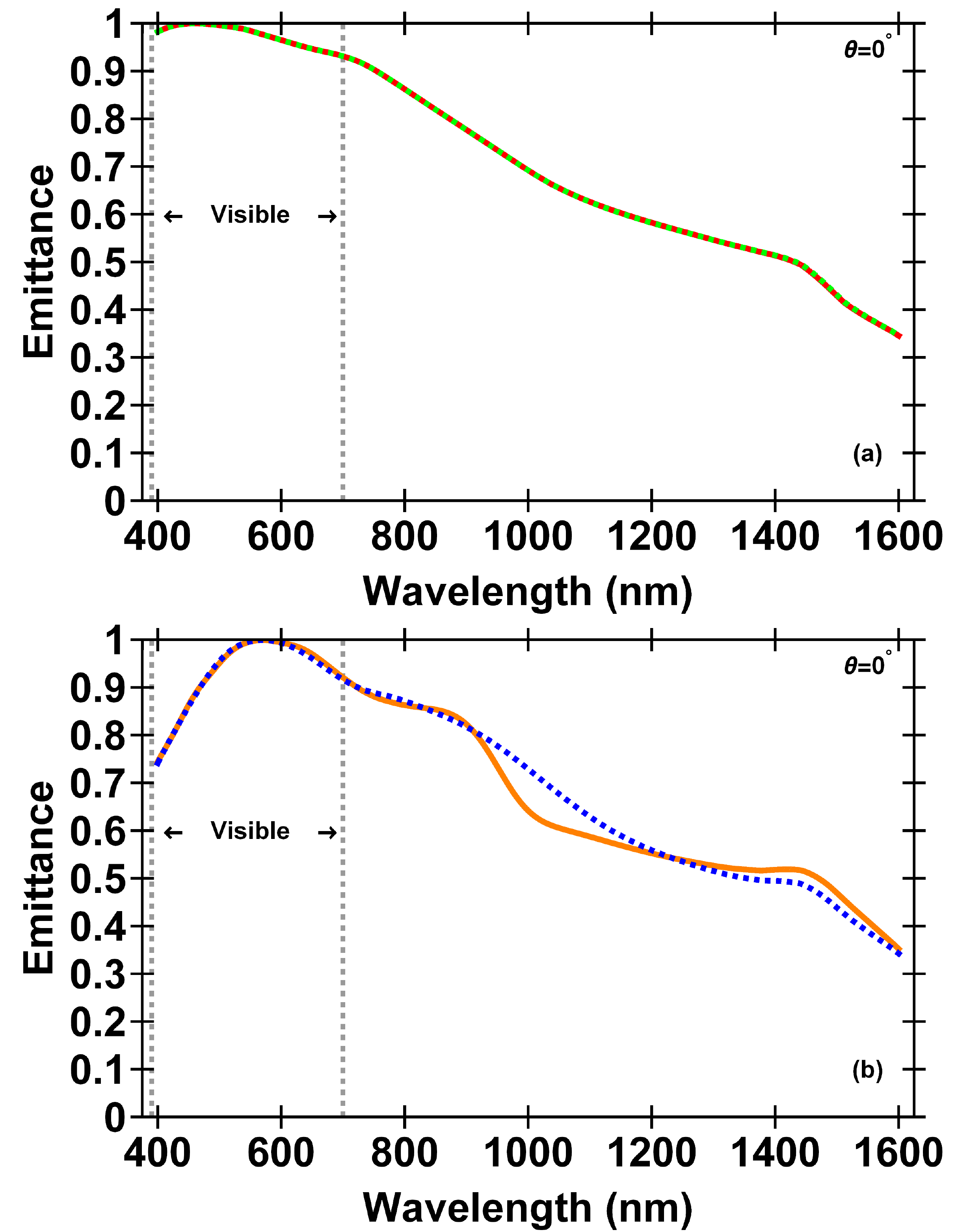}\\
   \caption{(a) Comparison of the emittance spectra in the normal direction of the optimized TA structure described in Fig. 2 (solid line) and the reduced TA structure which consists only of two layers (dashed line): a thin tungsten layer and an air gap above a thick tungsten substrate (Fig. 7). The thicknesses of the two layers of the reduced TA structure are the same as the thicknesses of the first two layers of the optimized TA structure. (b) Comparison of the emittance spectra in the normal direction of the optimized TSiC structure described in Fig. 2 (solid line) to the reduced TSiC structure which consists of a silicon carbide layer above a thick tungsten substrate (dashed line). The thickness of the silicon carbide layer of the reduced TSiC structure is the same as the thickness of the first silicon carbide layer of the optimized TSiC structure.}
   \label{1}
 
 \end{figure}
\begin{figure}[!ht]
   \centering
   \includegraphics[width=6.5cm]{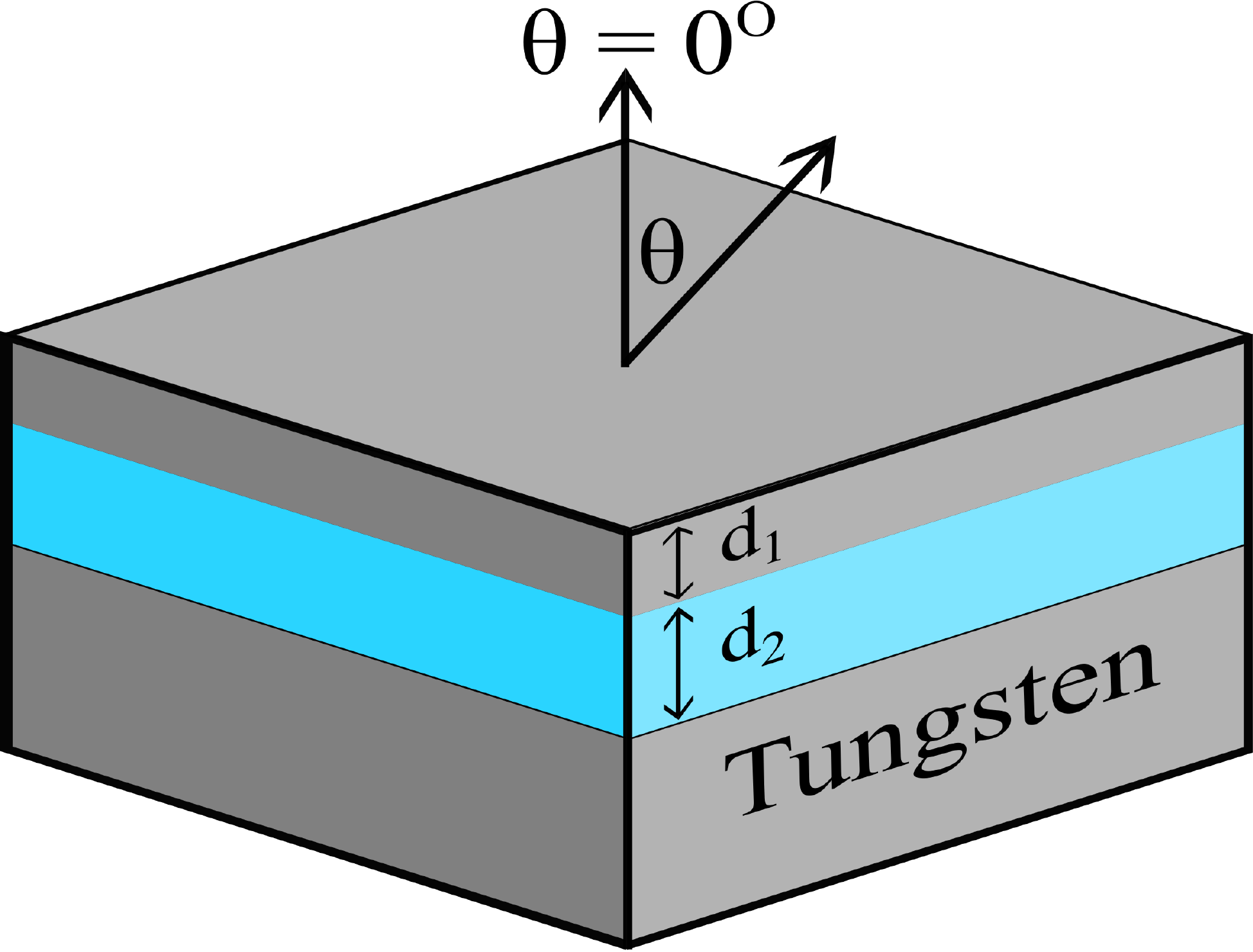}\\
   \caption{Schematic of the reduced structure. It consists of a single layer of tungsten and a dielectric layer above a semi-infinite tungsten substrate.}\label{1}   
 \end{figure}

In addition, we found that in the optimized TA structure the 500 $\rm nm$ thick air layer above the substrate [Fig. 4(a)] is not critical, since in a structure without this layer the emittance in the normal direction is smaller by no more than $\sim 0.5\%$ in the visible wavelength range [Fig. 6(a)]. Thus, the optimized TA structure is essentially equivalent to a 3.4 $\rm nm$ thick tungsten layer separated from the tungsten substrate by a 90 $\rm nm$ thick air gap. We will henceforth refer to a structure consisting of a thin tungsten layer and an air layer above a semi-infinite tungsten substrate (Fig. 7) as the reduced TA structure.

We note that for a thin metallic film placed $\lambda/4$ away from a perfect mirror, the required film thickness $d_1$ in order to achieve perfect impedance matching and therefore complete absorption in the film is \cite{subwavelengthstructures}
\begin{equation}
d_1 =\frac{\lambda}{4\pi n k},
\end{equation}
where $n$ and $k$ are the real and imaginary part of the refractive index of the metal, and $\lambda$ is the wavelength.
In the reduced TA structure, Fig. 6(a), the tungsten substrate is not a perfect mirror, and the thickness of the air gap above the substrate is not exactly $\lambda/4$ in the visible wavelength range. In addition, the optimized TA structure was obtained by optimizing the structure in the whole visible wavelength range rather than at a single wavelength. Despite these differences, we find that Eq. (8) approximately predicts the thickness of the thin tungsten layer in the optimized TA structure. More specifically, the required tungsten film thickness $d_1$ obtained from Eq. (8) in the visible wavelength range is 3.9 nm$ \le d_1 \le 5.5 $ nm, which is roughly the actual optimized layer thickness of 3.4 nm (Fig. 2). We note that, while the index of refraction of tungsten is wavelength-dependent and, thus, the optimum layer thickness is also wavelength-dependent, this 3.4 nm thickness choice for the thin tungsten layer leads to at least $73.5\%$ absorption in the tungsten layer in the whole visible wavelength range. It is also interesting to note that in the visible wavelength range, as the wavelength $\lambda$ increases, the $nk$ product for tungsten also increases \cite{indices}. Thus, the required tungsten film thickness $d_1$, for complete absorption obtained from Eq. (8), is relatively constant in the visible wavelength range. This suggests that tungsten is a uniquely suited material for thin film perfect absorbers in the visible.

   \begin{figure}[!ht]
     \centering
     \includegraphics[width=8.5cm]{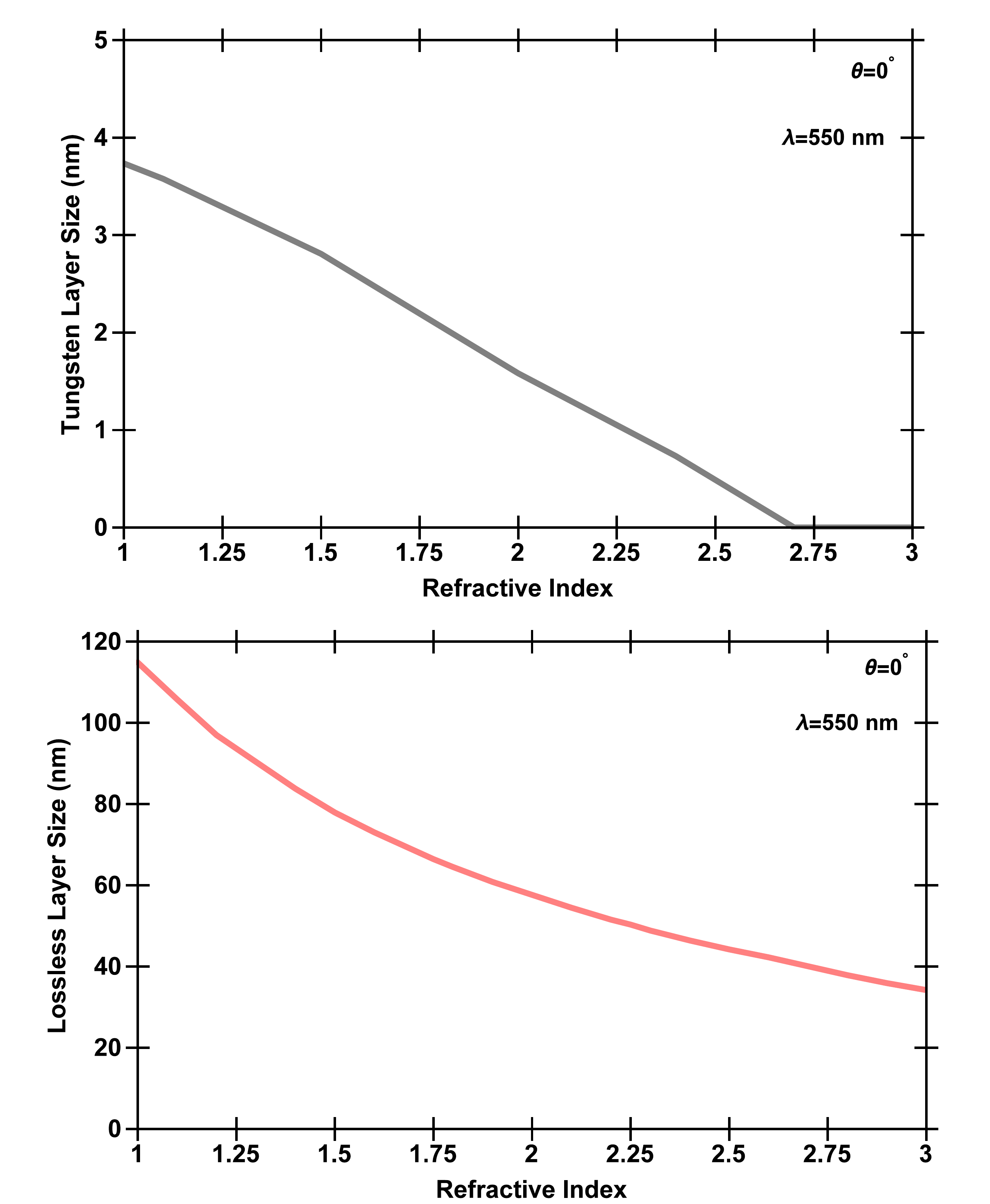}\\
     \caption{(a) Optimum thickness of the tungsten layer $d_1$ as a function of the refractive index of the dielectric layer $n$ for the structure of Fig. 7. The structure is optimized to maximize the emittance in the normal direction at the wavelength of $\lambda = 550 \rm \ nm$. (b) Optimum thickness of the dielectric layer $d_2$ as a function of the refractive index of the dielectric $n$ for the structure of Fig. 7. The structure is optimized to maximize the emittance in the normal direction at the wavelength of $\lambda = 550 \rm \ nm$.}
     \label{1}
   \end{figure}
   
In Fig. 4(b), we show the profile of the electric field amplitude normalized with respect to the field amplitude of the incident plane wave for the optimized TSiC structure. The structure is excited by a normally incident plane wave at the center of the visible spectrum, $ \lambda = 550\  \rm nm$. We observe that the optimized TSiC structure is nearly perfectly impedance-matched to air, as only $\sim 10\% $ of the incident power is reflected. In addition, we note that there is no field enhancement in the structure and conclude that, similar to the TA structures, the high emittance is not associated with strong resonances which is consistent with the broad-angle and broadband emittance of the structure [Fig. 3(e)]. 

However, we note that, compared to the TA structures, there are differences in the mechanisms which cause high emittance over the visible wavelength range. In contrast to the TA structures, in which the optimized structure possessed a thin tungsten layer which absorbed an overwhelming majority of incident light in the visible wavelength range, the tungsten layer thickness which borders air in the TSiC structure was optimized to be zero (Fig. 2). As a result, in the TSiC case, Eq. (8) is irrelevant. In Fig. 5(c), we show the emittance spectra of the TSiC structure as a function of the thickness of the first silicon carbide layer adjacent to air. As in the TA case, at a given wavelength the emittance as a function of the silicon carbide layer thickness exhibits peaks associated with Fabry-Perot resonances of the structure [Fig. 5(c)]. As before, the optimum layer thickness is selected based on the fact that a structure with a smaller thickness possesses smaller emittance in the visible; while, a structure with a larger thickness decreases the efficiency by emitting more in the infrared [Fig. 5(c)]. In addition, as in the TA case, the enhancement factor $Q$ as a function of the silicon carbide layer thickness exhibits additional peaks, but the maximum enhancement is obtained at the first peak, since for higher order peaks the emittance is not high in the entire visible wavelength range [Fig. 5(c)]. 

We also investigated the effect of changing the thickness of the first tungsten layer (below the first silicon carbide layer adjacent to air), and found that, as long as this layer thickness is larger than 60 nm, the absorption of the structure is unchanged. Based on this, we investigated the properties of a reduced TSiC structure which consists of a single silicon carbide layer above a thick tungsten substrate. The thickness of the silicon carbide layer of the reduced TSiC structure is the same as the thickness of the first silicon carbide layer of the optimized three-layer TSiC structure (39 nm). We found that for the reduced TSiC structure the emittance in the normal direction is smaller by no more than $\sim 0.5\%$ in the visible wavelength range compared to the optimized TSiC structure [Fig. 6(b)]. We note, however, that the emittance changes by approximately $10\%$ in some regions of the near infrared wavelength range [Fig. 6(b)].

A single silicon carbide layer can lead to high emittance in the whole visible wavelength range, because such a layer can provide near-perfect impedance matching between air and the tungsten substrate. More specifically, the impedance $\eta_{\rm TSiC}$ of the reduced TSiC structure which consists of a silicon carbide layer above a thick tungsten substrate is
\begin{equation}
\eta_{\rm TSiC}=\eta_{\rm SiC}\frac{\eta_{\rm W}+i \eta_{\rm SiC} \tan \beta_{\rm SiC }d_{\rm 2}}{\eta_{\rm SiC}+i \eta_{\rm W} \tan \beta_{\rm SiC }d_{\rm 2}},
\end{equation}
where $\eta_{\rm W}$ and $\eta_{\rm SiC}$ are the impedance of tungsten and silicon carbide, respectively, 
$\beta_{\rm SiC}=2\pi n_{\rm SiC}/\lambda$, $n_{\rm SiC}$ is the refractive index of silicon carbide, and $d_{\rm 2}$ is the thickness of the silicon carbide layer. As an example, using Eq. (9) for 
$\lambda=570\  {\rm nm}$, we find $\eta_{\rm TSiC}=356.3 + 3.4i \ \Omega$ which is very close to the air impedance $\eta_{\rm air}=377 \ \Omega$.

As mentioned above, we found that the optimized TA structure (Fig. 2) is essentially equivalent to the reduced TA structure consisting of a thin tungsten layer and an air layer above a thick tungsten substrate (Fig. 7). Similarly, the optimized TSiC structure (Fig. 2) is essentially equivalent to a reduced TSiC structure which consists of a single silicon carbide layer above a thick tungsten substrate. In an effort to provide a unified description of the TA and TSiC structures, we investigate the optimum thickness of the tungsten and dielectric layers as a function of the refractive index of the dielectric layer $n$ for the structure of Fig. 7. For simplicity, the structure is optimized to maximize the emittance at a single wavelength of $\lambda = 550 \rm \ nm$ near the center of the visible wavelength range. For $n=1$ (air), the optimum thickness of the tungsten layer was found to be 3.9 nm. Note that this thickness is slightly different than the optimized value of 3.4 nm in Fig. 2, since here we optimize the structure at a single wavelength rather than in the whole visible wavelength range. As $n$ is increased, the optimum tungsten layer thickness decreases, and becomes zero for $n \ge 2.6$ [Fig. 8(a)]. Thus, for lower refractive indices $n$, the presence of a thin tungsten layer which absorbs most of the incident light is beneficial. For higher refractive indices, the thin tungsten layer is not required, since the single dielectric layer can provide near-perfect impedance matching between air and the tungsten substrate. In this case, the incident light is mostly absorbed in the tungsten substrate. Silicon carbide has an index of refraction of $2.62\le n \le 2.77$ over the visible wavelength range, and thus a single silicon carbide layer can provide near-perfect impedance matching without the presence of a thin metallic absorbing layer. For the dielectric layer, as expected, as the refractive index $n$ increases, the optimum thickness of the layer decreases [Fig. 8(b)].

\section{Conclusion}

In this paper, we designed broadband thermal emitters in the visible wavelength range for use as lightbulb filaments. We considered structures composed of alternating layers of tungsten and air (TA structures) or tungsten and silicon carbide (TSiC structures) of varying aperiodic thicknesses above a semi-infinite tungsten substrate. We used a hybrid optimization method, consisting of a micro-genetic global optimization algorithm coupled to a local optimization algorithm, to maximize the power emitted by the structures in the visible. We then maximized the efficiency of the structures without significantly decreasing the power emitted in the visible. 
  
Both the TA and TSiC optimized structures that we obtained using this process show drastically improved emittance over the visible wavelength range when compared to bulk tungsten. The emittance of the optimized structures approaches the one of a perfect blackbody. The optimized structures are nearly perfectly impedance-matched to air. Additionally, there is no field enhancement in the structures, and therefore the high emittance is not associated with any strong resonances, which is consistent with the broad-angle and broadband emittance of the structures.

We also found that the optimized TA structure is essentially equivalent to a reduced TA structure consisting of a thin tungsten layer, and an air layer above a thick tungsten substrate. Similarly, we found that the optimized TSiC structure is essentially equivalent to a reduced TSiC structure which consists of a single silicon carbide layer above a thick tungsten substrate. In this case, the single silicon carbide layer leads to high emittance in the whole visible wavelength range, because it provides near-perfect impedance matching between air and the tungsten substrate.  

To implement our proposed lighting scheme we propose manufacturing sheets of the material inside of a transparent chamber filled with an inert gas (or evacuated) with current running through the tungsten substrate to provide the heating needed for thermal emission. To replace an incandescent light bulb we propose simply taking the usual solid tungsten filament, which is a cylinder with a $\sim 5\  \mu$m diameter and coating it with the multi-layered structure proposed here. This can be done by vapor deposition or molecular beam epitaxy. Given that the layers are on the scale of nanometers, the curvature of the filament is negligible; therefore, the planar 1D calculation still holds quantitatively. To reduce evaporation effects, one might wish to reduce the surface area to volume ratio and go to 2D or 3D structures (where a TA structure is feasible to fabricate via, for example, ion etching or a modified silicon layering process) \cite{3dpc}. More work needs to be done to model and optimize such structures, but our results in this work indicate qualitatively, via the John-Wang model, that similar emissivity (or better) is possible. In the John-Wang model, one assumes a 3D structure that has a perfectly spherical Brillouin zone that is the same for both the TE and TM modes. In such a model the structure is effectively 1D and our results here should hold qualitatively \cite{johnwang,johnwang2} .

This research was supported by the National Science Foundation (Award Nos. 1102301, 1254934, 1263236, 0968895), and a Fund for Innovation in Engineering Research (FIER) grant from the Louisiana State University College of Engineering. Jonathan P. Dowling wishes to also acknowledge support from the Air Force Office of Scientific Research and the Army Research Office.



\end{document}